\documentstyle{article}

\def\abstract#1{\vskip 7mm 
        \begin{center}{\large Abstract}\par \smallskip
                \begin{minipage}[c]{12cm}
                        \small #1
                \end{minipage}
        \end{center}
}
\def\title#1{\begin{center}{\Large\bf #1}\end{center}}
\def\author#1{\vskip 5mm \begin{center}{#1}\end{center}}
\def\address#1{\begin{center}{\it #1}\end{center}}
\makeatletter
\@ifundefined{lesssim}{}{}
\@ifundefined{gtrsim}{}{}
\def\vereq#1#2{\lower3pt\vbox{\baselineskip1.5pt \lineskip1.5pt
\ialign{$\m@th#1\hfill##\hfil$\crcr#2\crcr\sim\crcr}}}
\makeatother

\begin{document}

\title{%
  Noncommutative gravity in three dimensions
  coupled to point-like sources
}
\author{%
  Kiyoshi Shiraishi\footnote{E-mail:shiraish@sci.yamaguchi-u.ac.jp}
}
\address{%
  Faculty of Science, Yamaguchi University, \\
  Yamaguchi-shi, Yamaguchi, 753--8512, Japan
}
\abstract{
Noncommutative gravity in three dimensions 
with no cosmological constant is reviewed. 
We find a solution which describes the presence of
a torsional source. }

\section{What does `noncommutative' mean?}

Please see \cite{DN} for a review of noncommutative field theory, 
if you want to study more.

Consider noncommutative coordinates, for example
\begin{equation}
\left[\,x, y\,\right]=i\,\theta\, ,
\end{equation}
where $\theta$ is a real constant.
Then the `uncertainty' lies between $x$ and $y$, namely,
\begin{equation}
\Delta x\Delta y\ge \theta\, ,
\end{equation}
(where a numerical factor has omitted).
This means the existence of the minimal length scale.

If complex combinations of the coordinates,
$z=x+iy$ and $z=x-iy$, are introduced, they satisfy
\begin{equation}
\left[\,z, \bar{z}\,\right]=2\theta\, .
\end{equation}

There are different representations to describe the noncommutativity;
the commutative coordinate formalism with the star product, the Fock space
(operator) formalism, {\it etc.} (see \cite{DN} for details). In this talk, we
simply use them identically, unless the identification leads to confusions
(thus, we do not use
$\star$ in this talk). For example, we denote the equalities
\begin{equation}
1=\sum_{n=0}^{\infty}|n\rangle\langle n|\, ,\qquad
z=\sqrt{2\theta}\sum_{n=0}^{\infty}\sqrt{n+1}|n\rangle\langle n+1|\, ,\qquad
\bar{z}=\sqrt{2\theta}\sum_{n=0}^{\infty}\sqrt{n+1}|n+1\rangle\langle n|\, ,
\end{equation}
where ket and bra satisfy $z|0\rangle=0$, 
$z|n\rangle=\sqrt{2\theta}\sqrt{n}|n-1\rangle$,
$\bar{z}|n\rangle=\sqrt{2\theta}\sqrt{n+1}|n+1\rangle$, and so on.

Another example is
\begin{equation}
2\, (-1)^m 
L_m(2{r}^2/{\theta})\, e^{-r^2/\theta}=
|m\rangle\langle m|\, ,
\end{equation}
where $r^2=x^2+y^2$ and $L_n(x)$ is the Lagurre polynomial.

For later use, we {\it define}
\footnote{Note that here ${\tt \frac{1}{z}}$ is defined as operator
formalism and this may differ from $z^{-1}$ in the usual star product 
formalism.} 
the inverse of $z$, $\bar{z}$ as
\begin{equation}
{\tt
\frac{1}{z}}\equiv\frac{1}{\sqrt{2\theta}}\sum_{n=0}^{\infty}\frac{1}{\sqrt{n+1}}
|n+1\rangle\langle n|\, ,\quad
{\tt \frac{1}{\bar{z}}}\equiv\frac{1}{\sqrt{2\theta}}\sum_{n=0}^{\infty}
\frac{1}{\sqrt{n+1}}|n\rangle\langle n+1|\, .
\end{equation}

This definition leads to $z{\tt \frac{1}{z}}={\tt
\frac{1}{\bar{z}}}\bar{z}=1$, however, 
\begin{equation}
{\tt \frac{1}{z}} z=\bar{z}{\tt \frac{1}{\bar{z}}}=1-|0\rangle\langle 0|\, .
\end{equation}

Thus the derivative of ${\tt \frac{1}{z}}$ and ${\tt \frac{1}{\bar{z}}}$ is
\begin{equation}
\partial_{\bar{z}}{\tt \frac{1}{z}}=
\frac{1}{2\theta}\left[z,{\tt \frac{1}{z}}\right]=
\frac{1}{2\theta}|0\rangle\langle 0|=
\frac{1}{\theta}e^{-r^2/\theta}\, ,\quad 
\partial_{z}{\tt \frac{1}{\bar{z}}}=
\frac{1}{2\theta}\left[{\tt \frac{1}{\bar{z}}}, \bar{z}\right]=
\frac{1}{2\theta}|0\rangle\langle 0|=
\frac{1}{\theta}e^{-r^2/\theta}\, .
\end{equation}

Interestingly enough, in the commutative limit, we find
\begin{equation}
\frac{1}{\theta}e^{-r^2/\theta}\stackrel{\theta\rightarrow 0}{\longrightarrow}
\pi\, \delta(x)\delta(y)\, .
\end{equation}

\section{Why `noncommutative'?}

We expect that noncommutative field theory is worth studying,
because:

\begin{itemize}
\item the fundamental minimal length $\sqrt{\theta}$ may
avoid singularities and infinities which arise in usual
field theories.
\item the effective field theory of string or brane or M-theory in the
presence of the background $B$ field can be expressed as a noncommutative
field theory.
\item the analysis of quantum Hall systems with a constant magnetic field
admits the formulation with noncommuting coordinates.
\end{itemize}

\section{Three dimensional noncommutative gravity}
\dots has a long history.

In this talk, we concentrate our attention on noncommutative gravity 
in three dimensions.
\footnote{See Nair\cite{Nair} for an extension to the other dimensions.}

Three dimensional Chern-Simons noncommutative gravity was studied
by Ba\~nados {\it et al.}\cite{BCGSS} and 
more recently by Cacciatori {\it et al.}\cite{CKMZ}.

We would like to study noncommutative gravity in three dimensions
with no cosmological constant. Further, we wish to find exact
solutions whose spatial coordinates are mutually noncommutative.

The signature is taken to be Euclidean, and the coordinates are 
denoted as
\begin{equation}
x^1=x\, ,\quad
x^2=y\, ,\quad
x^3=\tau\, ,\qquad
{\rm where}\quad \left[\,x, y\,\right]=i\,\theta.
\end{equation}

We define a matrix-valued dreibein one-form and a connection one-form as
\begin{equation}
e=e^a J_a+e^4 i\, ,\quad \omega=\omega^a J_a+\omega^4 i\, ,
\quad{\rm where}\quad
J_1=\frac{i}{2}\sigma_1,\quad
J_2=-\frac{i}{2}\sigma_2,\quad
J_3=\frac{i}{2}\sigma_3\, .
\end{equation}
A matrix-valued torsion two-form and a curvature two-form
are given by
\begin{equation}
{\cal T}=de+\omega\wedge e+e\wedge\omega\, ,
\qquad {\cal R}=d\omega+\omega\wedge\omega\, .
\end{equation}
The vacuum solution of `Einstein equation' satisfies
\begin{equation}
{\cal R}={\cal T}=0\, .
\end{equation}
Unless the Abelian field $e^4$ and $\omega^4$ vanish,
we cannot regard this model as that for noncommutative gravity with 
an arbitrary value of $\theta$.

\section{Difficulties in noncommutative gravity I}

If we choose the spin connection as
\begin{equation}
\omega=\frac{\alpha}{2}\left({\tt \frac{1}{z}}dz-{\tt \frac{1}{\bar{z}}}
d\bar{z}\right)\left(
\begin{array}{rr}
1 & 0 \\
0 & -1
\end{array}
\right)\, ,
\end{equation}
we obtain the curvature
\begin{equation}
{\cal R}=\frac{\alpha}{2\theta}|0\rangle\langle 0|dz\wedge d\bar{z}
\left(
\begin{array}{rr}
1 & 0 \\
0 & -1
\end{array}
\right)+\frac{\alpha^2}{4}
\left[{\tt \frac{1}{z}}, {\tt \frac{1}{\bar{z}}}\right]dz\wedge d\bar{z}
\left(
\begin{array}{rr}
1 & 0 \\
0 & 1
\end{array}
\right)\, .
\end{equation}

In the commutative limit, this can be regarded as the
curvature of the spacetime where the point particle with mass
$-\alpha/(8G)$ is located at the origin (where $G$ is the Newton constant)
\cite{DJT}. 
For an arbitrary value of $\theta$, however, the Abelian part remains;
its interpretation is difficult, as for a theory of gravity.

\section{Difficulties in noncommutative gravity II}

In the matrix form, the local Lorentz transformation can be expressed as
\begin{equation}
e'=U^{-1}e\,U\, ,\qquad\omega'=U^{-1}\omega\, U+U^{-1}dU\, ,
\end{equation}
where $UU^{-1}=1$.

Under these transformation, ${\cal T}$ and ${\cal R}$ becomes
\begin{equation}
{\cal T}'=U^{-1}{\cal T}U\, ,\qquad {\cal R}'=U^{-1}{\cal R}U\, .
\end{equation}
Then the 
equations of motion ${\cal T}={\cal R}=0$ is unchanged.

On the other hand, the local translations can be written as
\begin{equation}
e'=e+d\rho+\omega\,\rho-\rho\,\omega\, ,\qquad
\omega'=\omega\, ,
\end{equation}
Under these transformation, ${\cal T}$ and ${\cal R}$ becomes
\begin{equation}
{\cal T}'={\cal T}+{\cal R}\rho-\rho {\cal R}\, ,\qquad {\cal R}'={\cal R}\, .
\end{equation}
The equations of motion ${\cal T}={\cal R}=0$ is unchanged
also by the translation.

Therefore, we can construct vacuum solutions as pure gauge.
The matrices which satisfy $UU^{-1}=1$ take the form
(modulo rigid rotations)
\begin{equation}
U=\left(
\begin{array}{cc}
S & 0 \\
P & S^{-1}
\end{array}
\right)\, ,\qquad
U^{-1}=\left(
\begin{array}{cc}
S^{-1} & P \\
0 & S
\end{array}
\right)\, ,
\end{equation}
where $SS^{-1}=1$, $S^{-1}S=1-P$, $P^2=P$, and
$SP=PS^{-1}=0$ are required.

Then the pure-gauge connection is
\begin{equation}
\omega_g=U^{-1}dU=
\left(
\begin{array}{cc}
S^{-1} & P \\
0 & S
\end{array}
\right)
\left(
\begin{array}{cc}
dS & 0 \\
dP & dS^{-1}
\end{array}
\right)=
\left(
\begin{array}{cc}
S^{-1}dS+PdP & PdS^{-1} \\
SdP & SdS^{-1}
\end{array}
\right)\, ,
\end{equation}

Unfortunately, this includes the Abelian part in general
(${\rm Tr}\,\omega$).
Thus the interpretation of this type of solutions is also
unclear in a theory of gravity.

\section{A solution with a torsional source}

Now we choose
\begin{equation}
e=\frac{i}{2}\left\{d\tau+\frac{GS}{2i}
\left({\tt \frac{1}{z}}dz-{\tt \frac{1}{\bar{z}}}
d\bar{z}\right)\right\}
\left(
\begin{array}{rr}
1 & 0 \\
0 & -1
\end{array}
\right)+\frac{i}{2}
\left(
\begin{array}{rr}
0 & dz \\
d\bar{z} & 0
\end{array}
\right)\, ,
\end{equation}
where $S$ is a constant
and $\omega=0$.

Then we obtain
\begin{equation}
{\cal T}=\frac{GS}{4\,\theta}\,|0\rangle\langle 0|\,d\bar{z}\wedge dz
\left(
\begin{array}{rr}
1 & 0 \\
0 & -1
\end{array}
\right)\, ,\qquad
{\cal R}=0\, .
\end{equation}

In the commutative limit, this corresponds to the solution
obtained by Deser, Jackiw and `t Hooft \cite{DJT} in the case of
the mass of the point particle is zero.
For a finite $\theta$, the torsional source has a finite extension.

This one-body solution can be generalized to the $N$-body
solution,
\begin{equation}
e=\frac{i}{2}\left\{d\tau+\sum_{a=1}^N\frac{GS_a}{2i}
\left({\tt \frac{1}{z-z_a}}dz-{\tt \frac{1}{\bar{z}-\bar{z}_a}}
d\bar{z}\right)\right\}
\left(
\begin{array}{rr}
1 & 0 \\
0 & -1
\end{array}
\right)+\frac{i}{2}
\left(
\begin{array}{rr}
0 & dz \\
d\bar{z} & 0
\end{array}
\right)\, ,
\end{equation}
where ${\tt z_a}$, ${\tt \bar{z}_a}$ and $S_a$are constants and 
${\tt \frac{1}{z-z_a}}$ and ${\tt \frac{1}{\bar{z}-\bar{z}_a}}$ are defined by
\begin{equation}
{\tt \frac{1}{z-z_a}}=\sum_{m=0}^{\infty} {\tt z_a}^m\left({\tt
\frac{1}{z}}\right)^{m+1}\, ,
\qquad
{\tt \frac{1}{\bar{z}-\bar{z}_a}}=\sum_{m=0}^{\infty}
{\tt \bar{z}_a}^m\left({\tt \frac{1}{\bar{z}}}\right)^{m+1}\, .
\end{equation}

\section{Wave equation}

Now we can write down a wave equation for a massless scalar field around
one torsional source,
\begin{equation}
\left\{-\partial_t^2+2\left(D_zD_{\bar{z}}+D_{\bar{z}}D_z\right)\right\}
\phi=0\, ,
\end{equation}
where
\begin{equation}
D_z\equiv \partial_z-\frac{GS}{2i}{\tt \frac{1}{z}}\partial_t\, ,\qquad
D_{\bar{z}}\equiv
\partial_{\bar{z}}+\frac{GS}{2i}{\tt \frac{1}{\bar{z}}}\partial_t\, .
\end{equation}
Here we changed the signature into the Lorentzian one.

Solving the noncommutative differential equation,
we will analyze the wave scattering by the torsional source.
The scattering process is nontrivial for a small impact parameter
$\approx\sqrt{\theta}$.
The investigation of the wave scattering will be reported elsewhere\cite{fw}.

\section{Open problems}

\begin{itemize}
\item How can the Abelian fields in this formalism have certain meanings
in a theory of gravity?
\item How can we obtain `conical' ({\it i.e.} massive) solution?
\item How can we take global properties of spacetime into account?
How and when do we have to use a nonocommutative torus and sphere?
\end{itemize}



\end{document}